 \documentclass[smallabstract,smallcaptions]{dccpaper}

\usepackage{epsfig}
\usepackage{amsmath}
\usepackage{amssymb}
\usepackage{color}
\usepackage{url}
\usepackage{hyperref}
\usepackage{multirow}
\usepackage{multicol}
\usepackage{subcaption}
\newlength{\figurewidth}
\newlength{\smallfigurewidth}

\begin{document}

\title
{\large
\textbf{LFZip: Lossy compression of multivariate floating-point \\time series data via improved prediction}
}

\author{%
Shubham Chandak$^1$, Kedar Tatwawadi$^1$, Chengtao Wen$^2$,\\ Lingyun Wang$^2$, Juan Aparicio$^2$, Tsachy Weissman$^1$\\[0.5em]
{\small\begin{minipage}{\linewidth}\begin{center}
\begin{tabular}{ccc}
$^1$Stanford University, CA, USA & \hspace*{0.5in} & $^2$Siemens Corporation, USA\\
\end{tabular}
\end{center}\end{minipage}}\\
\{schandak, kedart\}@stanford.edu, chengtao.wen@siemens.com
}

\maketitle
\thispagestyle{empty}

\begin{abstract}
Time series data compression is emerging as an important problem with the growth in IoT devices and sensors. Due to the presence of noise in these datasets, lossy compression can often provide significant compression gains without impacting the performance of downstream applications. In this work, we propose an error-bounded lossy compressor, LFZip, for multivariate floating-point time series data that provides guaranteed reconstruction up to user-specified maximum absolute error. The compressor is based on the prediction-quantization-entropy coder framework and benefits from improved prediction using linear models and neural networks. We evaluate the compressor on several time series datasets where it outperforms the existing state-of-the-art error-bounded lossy compressors. The code and data are available at \url{https://github.com/shubhamchandak94/LFZip}.
\end{abstract}

\Section{Introduction}
With the rapid increase in smart machines, IoT devices and sensors collecting and transmitting large volumes of measurement data, it has become essential to consider data compression strategies to reduce the transmission volume. This work focuses on the problem of multivariate time series compression, which arises in several applications such as manufacturing processes, medical measurements, activity trackers, autonomous vehicles, power consumption etc. In a number of cases, the time series consists of high-frequency floating-point data. This data typically contains measurement noise due to which lossy compression can provide significantly better compression without adversely impacting the downstream applications. In some cases, lossy compression can lead to improvement in the performance of downstream applications due to implicit denoising of the data \cite{ochoa2016effect}. In several of these scenarios, the compression is performed on an edge device that collects the data and transmits it to the cloud. These edge devices are usually computation and communication bandwidth constrained and hence the compression solutions must be real time and suitable for these devices.


A lossy compressor consists of an encoder and a decoder, where the encoder compresses the original time series $x_1, x_2,\dots,x_n$ to obtain the compressed bit stream. This compressed bit stream can then be decompressed by the decoder to obtain the reconstructed time series $\hat{x}_1, \hat{x}_2,\dots,\hat{x}_n$ which differs from the original time series to an extent acceptable for a specific application. The error in the reconstructed time series with respect to the original time series is measured in terms of a distortion function, such as the mean squared error, mean absolute error or maximum absolute error. The optimal distortion function is highly dependent on the downstream applications that work with the reconstructed time series, but in absence of specific domain knowledge, maximum absolute error, defined as $\max_{i=1,\dots,n} |x_i-\hat{x}_i|$, is a generally acceptable distortion measure. Note that a maximum absolute error of $\epsilon$ implies that the original and the reconstructed time series differ by at most $\epsilon$ at any time step.

\SubSection{Our Contributions}
    We propose LFZip (Lossy Floating-point Zip), a lossy compressor for time series data based on the prediction-quantization-entropy coder framework \cite{gersho2012vector} which achieves significant improvement in compression over the previous state-of-the-art compressors. LFZip works under the maximum absolute error distortion metric where the maximum allowable absolute error is a user-specified parameter. 
    LFZip also supports compression of multivariate time series where the dependencies across the variables are exploited to further boost the compression.
    
    LFZip is available as an open source tool, providing a easily extensible framework supporting several prediction models including linear predictors and neural networks.


\SubSection{Previous Work}
There have been several works on lossless and lossy compression of floating-point time series and multidimensional scientific data. For lossless compression of floating-point data, specialized compressors such as FPZIP \cite{fpzip} outperform general-purpose compressors on multidimensional datasets. Several lossy compressors have also been proposed, which allow much better compression at the expense of some acceptable level of distortion. We next discuss a few of the existing works in literature.

Swinging door \cite{bristol1990swinging} and Critical Aperture \cite{williams2006critical} algorithms retain only a subset of the points in the time series based on the maximum error constraint and use linear interpolation during decompression. These are widely used in certain domains \cite{evsystems,osisoft} due to their suitability for computationally constrained systems. Another line of work uses polynomial or regression models to predict the next point in the time series and then quantizes the prediction error. Compressors in this category include SZ \cite{sz1,sz2,sz3,sz4}, ISABELA \cite{lakshminarasimhan2013isabela} and NUMARCK \cite{chen2014numarck}, of which SZ is the state-of-the-art compressor under maximum error distortion. Finally, some of the compressors \cite{hawkins2012algorithm,sasaki2015exploration} apply a transform to the original time series and then perform quantization in the transformed domain. 
Note that some of these compressors are not specific to time series and also support multidimensional scientific datasets. We refer the reader to \cite{sz1} for a more detailed survey on lossy compressors under different distortion measures.

We should note that the general approach of prediction-quantization-entropy coding employed in this work has been extensively applied for lossy compression in the context of speech \cite{o1988linear}, images \cite{webp} and videos \cite{feller2011vp8}, where domain-specific prediction models and distortion measures are used, and has also been studied theoretically (see \cite{gersho2012vector}). While working within this traditional framework, this work attempts to utilize advances in prediction models to achieve improved error-bounded lossy compression of time series data. This follows a line of similar efforts to improve lossless data compression using powerful prediction models such as neural networks  \cite{goyal2019deepzip,liu2019decmac,pagin2017neural}. 

We next discuss the methods employed in the proposed compression framework.

\Section{Methods}
    
    


\SubSection{Encoding and Decoding Framework}

\begin{figure}[htbp]
	\centering
	\includegraphics[width=\textwidth]{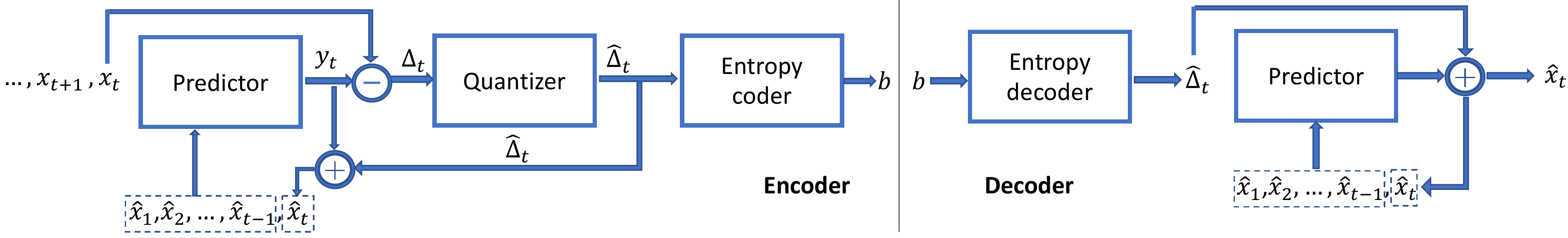}
	\caption{Encoder and decoder framework in LFZip.}
	\label{fig_encoder}
\end{figure}

The encoding and decoding framework for LFZip is summarized in Figure \ref{fig_encoder}. The encoder processes the input one symbol at a time, and consists of 3 stages: predictor, quantizer and entropy coder. For simplicity, we describe these in detail for the case of univariate time series. In case of multivariate time series, only the prediction step is modified while the quantizer and entropy coder act independently on each variable. 


\vspace{0.05in}
    \noindent
    \textbf{Predictor:} At every time-step $t$, the predictor block tries to guess the value $x_t$, based on the past reconstructions. Formally, the predictor block fits a function $P_t(\hat{x}_1, \hat{x}_2, \ldots, \hat{x}_{t-1})$, to obtain the prediction $y_t$. Note that the predictor is restricted to be causal in nature and is applied on the reconstructed values (rather than the input) so that the same procedure can be applied during the decompression. The predictor function $P_t$ itself can be adaptive but must satisfy the causality constraint and should be trained only on the reconstructed values. More details about the predictors used in the framework and the training procedure are presented later.
    
    For multivariate time series with $k$ variables per time step, the prediction of time step $t$ for the $j^{\textrm{th}}$ variable can be based on not only the time steps $1,\dots,t-1$ for the $j^{\textrm{th}}$ variable, but also on time steps $1,\dots,t-1$ for the other variables and time step $t$ for the first $j-1$ variables (to ensure causality). This allows the predictor to exploit dependencies across variables and provides improved compression in some cases. 

    \vspace{0.05in}
    \noindent\textbf{Quantizer:} The floating-point prediction error is quantized in this step so as to satisfy the maximum absolute error constraint. 
    Let $\Delta_t = x_t - y_t$ be the difference between the true value and the predictor output at time step $t$. The quantizer performs uniform scalar quantization of this $\Delta_t$ using a step size of $2\epsilon$ to obtain the quantized output $\hat{\Delta}_t$. The final reconstructed output $\hat{x}_t$ is then given by $\hat{x}_t = y_t + \hat{\Delta}_t$. Note that the uniform quantization guarantees that $|\hat{\Delta}_t - \Delta_t| \leq \epsilon$, which in turn implies that $|\hat{x}_t - x_t| \leq \epsilon$. We use 16-bit quantized output allowing for 65535 quantization bins (1 bin reserved for outliers as discussed below). We also tested 8-bit quantization, but the impact was negligible since the entropy coding stage is able to remove any redundancy introduced by 16-bit quantization.
    
    In practice, the presence of outliers or sudden change in statistics of the time series can lead to a large $\Delta_t$ value that lies outside the quantization range. In such cases, the reserved quantization bin is used and the data point $x_t$ is stored as a floating-point number, with the reconstruction $\hat{x}_t = x_t$.
    
    
    
    \vspace{0.05in}
    \noindent\textbf{Entropy Coder:} The final stage of the compression involves applying a universal lossless compressor on the quantized time series of the differences: $\hat{\Delta}_1, \hat{\Delta}_2, \ldots, \hat{\Delta}_n$ to obtain the variable length bit-stream $b$. LFZip uses BSC \cite{bsc}, an efficient BWT-based compressor, for this stage.

The Decoder operations are symmetric but occur in the reverse order (see Figure \ref{fig_encoder}). First the entropy-coded bit-stream $b$ is decoded to obtain $\hat{\Delta}_1, \hat{\Delta}_2, \ldots, \hat{\Delta}_n$. The reconstruction then occurs one time-step at a time, in a causal fashion. At each time-step, the predictor $P_t(\hat{x}_1, \hat{x}_2, \ldots, \hat{x}_{t-1})$ causally outputs the prediction $y_t$, which can then be used to output the reconstruction $\hat{x}_t$. Note that the same predictor function $P_t$ and adaptive training procedure must be used during the encoding and decoding.

\SubSection{Predictors}
Several prediction models can be utilized in the framework described. We look here at two classes of predictors of varying complexity supported by LFZip. 

\vspace{0.05in}
    \noindent \textbf{Normalized Least Mean Square Predictor (NLMS):} The Normalized Least Mean Square Predictor (NLMS) can be thought of as an adaptive linear prediction filter of window size $k$ ($32$ by default). The parameters of the linear filter are initialized with a fixed value and are updated at each time-step based on the mean square prediction error. The update procedure is similar to stochastic gradient descent, where the gradients are normalized before update. As the predictor contains very few parameters, we observed that in practice it requires no pre-training and adapts very quickly to changing input statistics. In practice, the NLMS predictor works well on various types of inputs and is the default predictor in the current implementation.

    
    
    
  
  \vspace{0.05in}
   \noindent \textbf{Neural Network based Predictors: } LFZip also supports more complex neural network based predictors. The current implementation supports different variants of the Fully Connected (FC) and the biGRU networks \cite{chollet2015keras} for univariate time series. The FC and biGRU networks take as input the window of past $k$ reconstruction symbols ($k=32$ by default) and output a floating-point prediction for $x_t$. As the NLMS predictor can be thought of as a single layer linear neural network, the FC and biGRU networks are strictly stronger models (in terms of expressibility). However, the larger number of parameters in FC and biGRU networks make them adapt much more slowly to the changing statistics in the time series. To resolve this issue, we employ offline training for neural network based predictors before the encoding step. 
   
      During the offline training, the model is trained on some given training data with early stopping performed with respect to validation data. The trained model is used as the predictor during compression, and the parameters can be optionally updated online during the compression. The utility of offline training highly depends upon on the similarity of the training data with the test data being compressed. 

   Note that while the training is performed on the true unquantized values, during the compression, the model performs prediction based on the quantized values. This can lead to worse performance in the case where the maximum error threshold $\epsilon$ is large. To resolve this issue, we approximately emulate the quantization process during the training by adding some appropriate noise to the inputs (based on the maximum error parameter $\epsilon$). We observed that adding noise during the training leads to 5-10\% improvement in compression. 
   We experimented with uniform and Gaussian noise models and observed that uniformly distributed noise model typically works better for the maximum error constraint. 


\Section{Results}
\begin{table}[htbp]
\begin{center}
{
\renewcommand{\baselinestretch}{1}\footnotesize
\resizebox{\textwidth}{!}{
\begin{tabular}{cclc}
\hline
\multirow{2}{*}{Name} & \multirow{2}{*}{Length} & \multirow{2}{*}{Description}  & BSC lossless\\ &&&compression ratio\\
\hline
\textit{acc} & 3.54M & Heterogeneity Activity Recognition - smartwatch accelerometer \cite{activity_recognition}&2.84\\
\textit{gyr} & 3.21M & Heterogeneity Activity Recognition - smartwatch gyroscope \cite{activity_recognition}&2.79\\
\textit{pow} & 2.05M & Household electric power consumption - active power \cite{household_power_consumption}&5.21\\
\textit{ppg} & 0.50M & Blood volume pulse/photoplethysmography (PPG) \cite{wesad}&2.48\\
\textit{gas} & 0.93M & Home activity monitoring - MOX gas sensors resistance \cite{ht_sensor_data}&4.97\\
\textit{dna} & 1.17M      & Nanopore DNA sequencing raw current data  &4.55\\
\textit{vib} & 1.55M      & Siemens healthy tool vibration data & 1.79\\
\textit{sen} & 0.75M      & Siemens sensor data & 4.27\\
\hline
\end{tabular}}}
\caption{\label{table:datasets}%
Datasets used for evaluation.}
\vspace{-0.5in}
\end{center}
\end{table}
\SubSection{Experimental setup}

We evaluated the proposed compressor LFZip on several time series datasets, spanning a variety of domains including smartwatch sensor data, household power consumption data, gas sensor array data, medical and genomic data, etc. as shown in Table \ref{table:datasets}. We also report results on two datasets obtained from Siemens. The first five datasets were chosen as a representative sample of the floating-point time series datasets from the UCI Machine Learning Repository \cite{uci}. Figure \ref{fig:plots} in the Appendix contains snapshots of all the datasets. The datasets can be accessed online on the LFZip GitHub repository (except for the Siemens datasets). 

We compare LFZip with two additional compressors SZ \cite{sz1,sz2,sz3,sz4} and critical aperture (CA) \cite{williams2006critical}. SZ is the current state-of-the-art compressor for maximum error distortion \cite{sz1}, while CA is widely used in the industry due to its low computational requirements \cite{evsystems,osisoft}. For SZ, we used the implementation available at \url{https://github.com/disheng222/SZ} (version: 2.1.7), while we implemented CA based on the description available at \cite{williams2006critical,evsystems,swingingdoor}. While certain implementations of CA do not use an entropy coding step, we apply BSC to the retained points for fair comparison with LFZip and SZ. Some of the datasets contained multivariate time series out of which a single variable was considered for fair comparison with CA and SZ as they don't natively support multivariate time series compression.\footnote{Note that although SZ supports multidimensional data compression, we found that this mode in fact leads to worse compression when run on multivariate time series, probably because SZ expects continuity along each dimension, which might not be true for the variables at a single time step.}  Results for multivariate time series compression using LFZip are discussed in a later section.




\SubSection{Results for LFZip (NLMS) for univariate time series data}
\begin{table}[htbp]
\begin{center}
{
\renewcommand{\baselinestretch}{1}\footnotesize
\resizebox{\textwidth}{!}{
\begin{tabular}{|c|c|ccc|}
\hline
\multirow{2}{*}{Dataset} & \multirow{2}{*}{Compressor} & \multicolumn{3}{c|}{Maximum error $\epsilon$}\\
& & $10^{-3}$ & $10^{-2}$ &  $10^{-1}$ \\
\hline
\multirow{3}{*}{\textit{acc}} &CA    & 2.84  & 3.01   & 5.19   \\
&SZ    & 3.25  & 5.05   & 11.00  \\
&LFZip (NLMS)  & \textbf{3.55}  & \textbf{5.86}   & \textbf{12.71}  \\
\hline
\multirow{3}{*}{\textit{gyr}} &CA   &2.88&	4.27&	10.75 \\
&SZ    & 4.26&	8.08&	24.79	 \\
&LFZip (NLMS)  &  \textbf{6.05}&	\textbf{12.26}&	\textbf{28.77}	\\
\hline
\multirow{3}{*}{\textit{pow}} &CA    & 5.05  & 6.23   & 12.47  \\
&SZ    & \textbf{5.09}  & \textbf{9.65}   & \textbf{23.99}  \\
&LFZip (NLMS)  & 4.17  & 7.37   & 17.98  \\
\hline
\multirow{3}{*}{\textit{ppg}} &CA    & 2.48  & 2.49   & 2.74   \\
&SZ    & 2.43  & 2.80   & 4.39   \\
&LFZip (NLMS)  & \textbf{3.18}  & \textbf{5.28}   & \textbf{9.13}   \\
\hline
\end{tabular}
\quad
\begin{tabular}{|c|c|ccc|}
\hline
\multirow{2}{*}{Dataset} & \multirow{2}{*}{Compressor} & \multicolumn{3}{c|}{Maximum error $\epsilon$}\\
& & $10^{-3}$ & $10^{-2}$ &  $10^{-1}$ \\
\hline
\multirow{3}{*}{\textit{gas}} &CA    & 16.97 & 64.36  & 245.51 \\
&SZ    & 22.69 & 75.84  & \textbf{299.65} \\
&LFZip (NLMS)  & \textbf{31.56} & \textbf{101.48} & 252.55\\
\hline
\multirow{3}{*}{\textit{dna}} &CA     & \textbf{4.54} &	4.54 &	4.86\\
&SZ    &4.03&	\textbf{4.55}&\textbf{8.62} \\
&LFZip (NLMS)  &3.04&	4.48&	8.40\\
\hline
\multirow{3}{*}{\textit{vib}} &CA    &  2.07&	4.85&	18.51  \\
&SZ    &  4.77&	11.77&	40.61\\
&LFZip (NLMS)  & \textbf{10.64} &	\textbf{22.36}&	\textbf{53.15}\\
\hline
\multirow{3}{*}{\textit{sen}} &CA   &4.34	&7.60	&125.04 \\
&SZ    & 6.55 &	20.58&	179.87 \\
&LFZip (NLMS)  & \textbf{6.88} &	\textbf{21.70} &	\textbf{180.98}\\
\hline
\end{tabular}}}
\caption{\label{table:main_results}%
Compression ratios for CA, SZ and LFZip (NLMS). Best results are boldfaced.}
\vspace{-0.25in}
\end{center}
\end{table}

Table \ref{table:main_results} shows the results for CA, SZ and LFZip (using NLMS predictor with default window size $k = 32$) for the eight univariate time series datasets and three values of the maximum error ($10^{-1}, 10^{-2}, 10^{-3}$). For comparison, Table \ref{table:datasets} shows the results for lossless compression with BSC, where BSC was chosen since it outperformed other lossless tools like Gzip, 7-zip, bzip2 and fpzip \cite{fpzip}. We see that lossy compression can lead to significant benefits over lossless compression, especially at higher maximum error constraints. From Table \ref{table:main_results}, we see that LFZip achieves the best compression in most cases, except for the \textit{pow} and \textit{dna} datasets. In general we found that the LFZip (NLMS) offers the most benefits for datasets that are difficult to approximate with piecewise linear functions or lower order polynomials (see Figure \ref{fig:plots} in the Appendix for snapshots of the datasets). The worse performance of LFZip (NLMS) on \textit{pow} and \textit{dna} datasets can be explained by the sudden jumps in the time series which are difficult to predict using a linear model. We will later see that we can overcome this limitation of NLMS predictors by using more powerful NN based predictors.

\SubSection{Results for LFZip (NLMS) for multivariate time series data}
\vspace{-0.1in}
\begin{table}[htbp]
\begin{center}
{
\renewcommand{\baselinestretch}{1}\footnotesize
\resizebox{\textwidth}{!}{
\begin{tabular}{|c|c|ccc|}
\hline
\multirow{2}{*}{Dataset} & \multirow{2}{*}{Mode} & \multicolumn{3}{c|}{Maximum error $\epsilon$}\\
& & $10^{-3}$ & $10^{-2}$ &  $10^{-1}$ \\
\hline
\textit{acc} &univariate    & 3.588	& 5.931	& 13.220   \\
(X, Y, Z)&multivariate    & 3.592 &	5.934	& 13.250  \\
\hline
\textit{gyr} &univariate    & 6.295& 13.605	& 34.181  \\
(X, Y, Z)&multivariate    & 6.409& 13.763 &	34.597 \\
\hline
\end{tabular}
\quad
\begin{tabular}{|c|c|ccc|}
\hline
\multirow{2}{*}{Dataset} & \multirow{2}{*}{Mode} & \multicolumn{3}{c|}{Maximum error $\epsilon$}\\
& & $10^{-3}$ & $10^{-2}$ &  $10^{-1}$ \\
\hline
\textit{gas} &univariate    & 26.239	& 63.304 &	152.378   \\
(8 sensors)&multivariate    & 27.614 &	75.179 & 204.006  \\
\hline
\textit{sen} &univariate    & 6.627	&   19.669	&   166.568   \\
(3 sensors)&multivariate    & 6.669 &  	20.334 &	304.878  \\
\hline
\end{tabular}}}
\caption{\label{table:multivariate_timeseries}%
LFZip (NLMS) compression ratios for multivariate time series (i) when each variable is compressed independently and (ii) when compressed together.}
\vspace{-0.25in}
\end{center}
\end{table}

LFZip can provide further improvement in compression for multivariate time series with dependencies across the variables. Table \ref{table:multivariate_timeseries} shows the results for \textit{acc} (3 variables: X, Y and Z), \textit{gyr} (3 variables: X, Y and Z), \textit{gas} (8 variables: different MOX gas sensors) and \textit{sen} (3 variables: different sensors) where the different variables are (i) compressed independently with LFZip in the univariate mode and (ii) compressed together with LFZip in the multivariate mode. We see significant gains of compressing the series together for the \textit{gas} and \textit{sen} datasets, but very minor improvement for \textit{acc} and \textit{gyr} datasets. The advantages for compressing multivariate time series together is likely to be most pronounced when the values for other variables can aid the predictor beyond the past values for the same variable. 

\SubSection{LFZip (NLMS) ablation experiments}
\begin{figure}[htbp]
	\centering
    \includegraphics[width=0.54\textwidth]{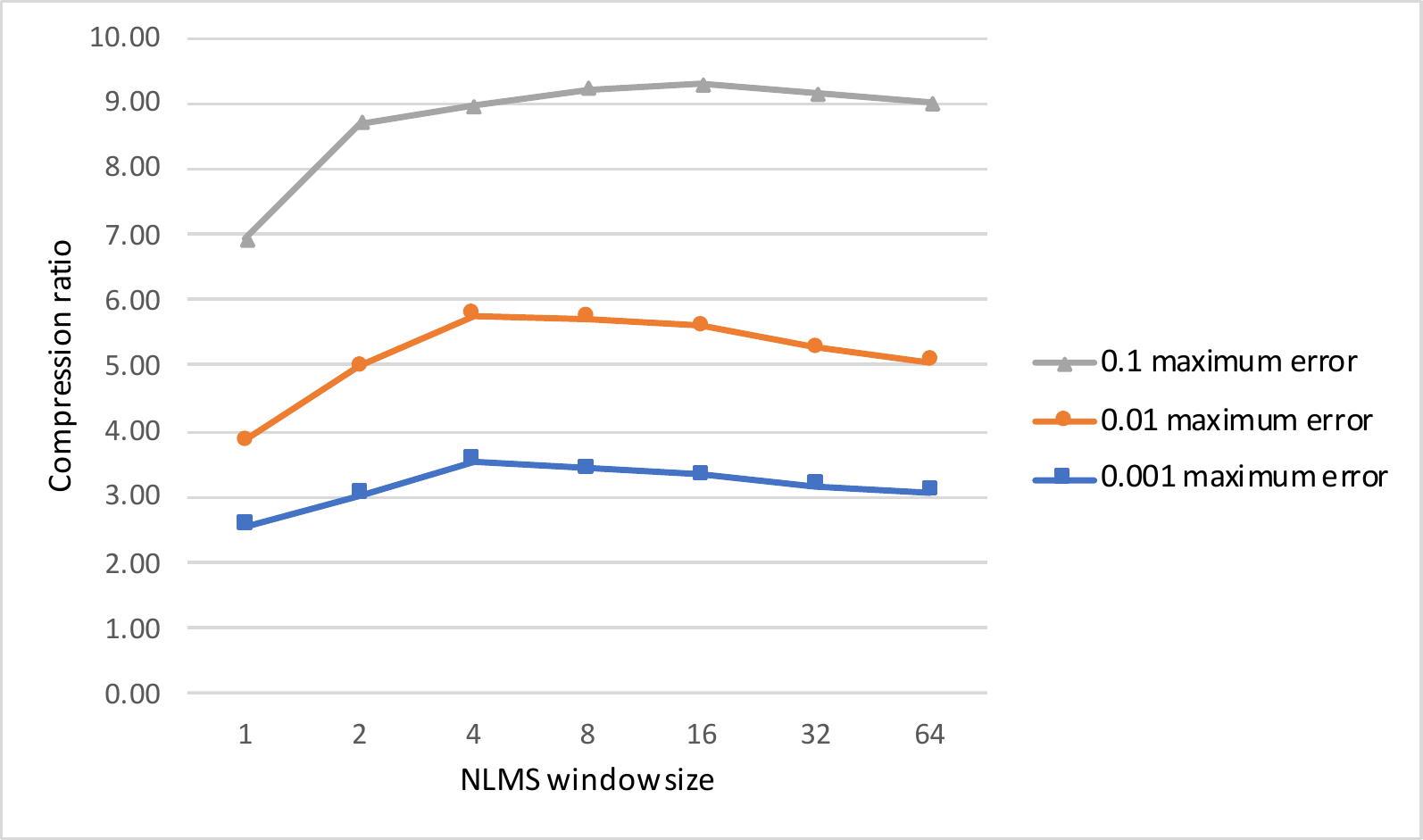}
  %
	\caption{Compression ratio for \textit{ppg} dataset as the NLMS window size is varied.}
	\vspace{-0.15in}
	\label{fig:nlms_parameters}
\end{figure}
In this section, we study the impact of the prediction and entropy coding stages for LFZip (NLMS). Figure \ref{fig:nlms_parameters} shows the impact of the NLMS window size ($k$) on the compression ratio for the \textit{ppg} dataset. We observe that the performance initially improves with increasing $k$ as the model becomes more expressive but at higher window sizes the performance becomes worse. This is likely because the higher number of parameters makes it slower at adapting to the changing statistics in the time series.

To understand the impact of the entropy coding stage on the performance of LFZip, we replaced the entropy coder BSC by 7-zip and Gzip for the \textit{ppg} dataset with maximum error $10^{-2}$. The compression ratios obtained were 5.28, 4.43 and 3.48 for BSC, 7-zip and Gzip, respectively. This shows that the entropy coder plays an important role in the framework and allows us to use a simple uniform scalar quantizer. We note that even with 7-zip as the entropy coder, LFZip achieves better compression than SZ for most datasets, showing the advantages of improved prediction.

\SubSection{Results for LFZip (NN) for univariate time series data}
\vspace{-0.1in}
\begin{table}[htbp]
\begin{center}
{
\renewcommand{\baselinestretch}{1}\footnotesize
\resizebox{0.9\textwidth}{!}{
\begin{tabular}{|c|c|cc|}
\hline
\multirow{2}{*}{Dataset} & \multirow{2}{*}{Compressor} & \multicolumn{2}{c|}{Maximum error $\epsilon$}\\
& & $10^{-2}$ &  $10^{-1}$ \\
\hline
\multirow{3}{*}{\textit{acc}} &SZ    &   4.64 &	9.38     \\
&LFZip (NLMS)    &   5.10 &	10.19   \\
&LFZip (NN)  &  \textbf{5.26} &	\textbf{10.78}  \\
\hline
\multirow{3}{*}{\textit{gyr}} &SZ    &   6.99 &	20.96  \\
&LFZip (NLMS)    &  10.22 &	23.33   \\
&LFZip (NN)  & \textbf{10.35}& \textbf{25.00} \\
\hline
\end{tabular}
\quad
\begin{tabular}{|c|c|cc|}
\hline
\multirow{2}{*}{Dataset} & \multirow{2}{*}{Compressor} & \multicolumn{2}{c|}{Maximum error $\epsilon$}\\
& & $10^{-2}$ &  $10^{-1}$ \\
\hline
\multirow{3}{*}{\textit{pow}} &SZ    &  \textbf{9.44} & 23.57    \\
&LFZip (NLMS)      &   7.21 &	17.74  \\
&LFZip (NN)   &  9.29 &	\textbf{25.38}    \\
\hline
\multirow{3}{*}{\textit{dna}} &SZ    &  4.45& 8.67  \\
&LFZip (NLMS)     & 4.46 & 8.40 \\
&LFZip (NN)   & \textbf{4.60} & \textbf{8.99} \\
\hline
\end{tabular}}}

\caption{\label{table:nn_results}%
Compression ratios for test datasets for SZ, LFZip (NLMS) and LFZip (NN). Best results are boldfaced.
}
\vspace{-0.25in}
\end{center}
\end{table}

Table \ref{table:nn_results} shows the results for LFZip with a neural network (NN) based predictor. For these experiments, we used a simple fully connected network with 4 hidden layers with 128 neurons each and first layer with 32 neurons, ReLU activation and batch normalization. Only the datasets for which the statistics were stationary over time were used for these experiments so as to allow offline training before compression. The datasets were divided into training, validation and test datasets of equal sizes, and the compression was performed only on the test dataset for all the compressors. For LFZip (NN), offline training was performed using the training and validation datasets for 5 epochs. During training, uniformly distributed noise in $[-0.05,0.05]$ was added to emulate the quantization noise. No adaptive training was employed during compression as that didn't seem to provide measurable benefits. 

From Table \ref{table:nn_results}, we see that LFZip (NN) improves the result over LFZip (NLMS) and SZ in all cases, except for \textit{pow} dataset with maximum error $10^{-2}$ where it is slightly worse than SZ. This shows that better prediction using stronger NN based models can lead to further improvement in compression for LFZip at the cost of increased computational complexity. 

\SubSection{Computational requirements}
The experiments were run on an Ubuntu 18.04 server with 2.2 GHz Intel Xeon processors and NVIDIA TITAN X Pascal GPUs. Note that the GPUs were used only during the training phase of the NNs, and the compression with both NLMS and NN predictors was performed on CPU with a single thread to ensure reproducibility and correct decompression. The current implementation of LFZip (NLMS) is written in Python and C++, and achieved a compression/decompression speed of $\sim$2M timesteps/s for univariate time series. This is about an order of magnitude slower than SZ but should be practical for most applications. LFZip (NN) is more computationally expensive with a speed of $\sim$1K timesteps/s for the model used above.

\Section{Conclusion}
We propose an error-bounded lossy compressor for multivariate floating-point time series based on the prediction-quantization-entropy coder framework. Using linear and neural network prediction models, the proposed compressor LFZip achieves higher compression ratios than the previous state-of-the-art compressors. Future work includes an optimized implementation for the neural network based framework, extension of the framework to multidimensional datasets and exploration of other predictive models to further boost compression.

\Section{References}
\bibliographystyle{IEEEbib}
\bibliography{refs}
\begin{figure}[htbp]
  \begin{subfigure}[b]{0.5\textwidth}
    \includegraphics[width=\textwidth]{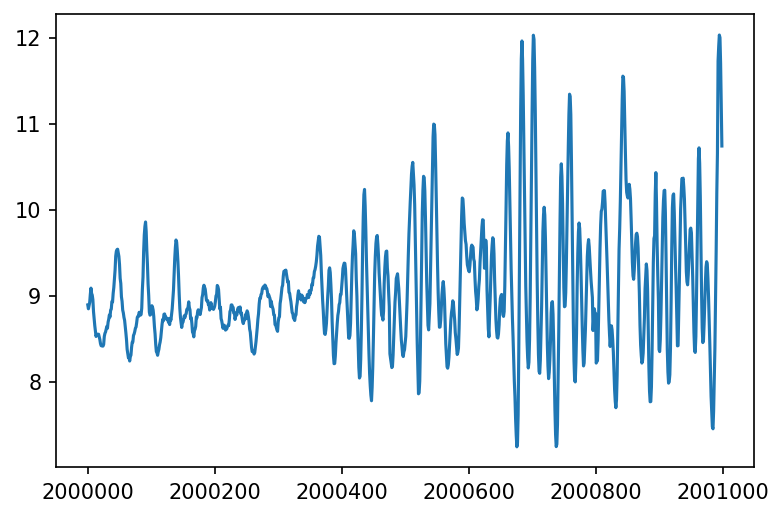}
    \caption{\textit{acc}}
  \end{subfigure}
    \begin{subfigure}[b]{0.5\textwidth}
    \includegraphics[width=\textwidth]{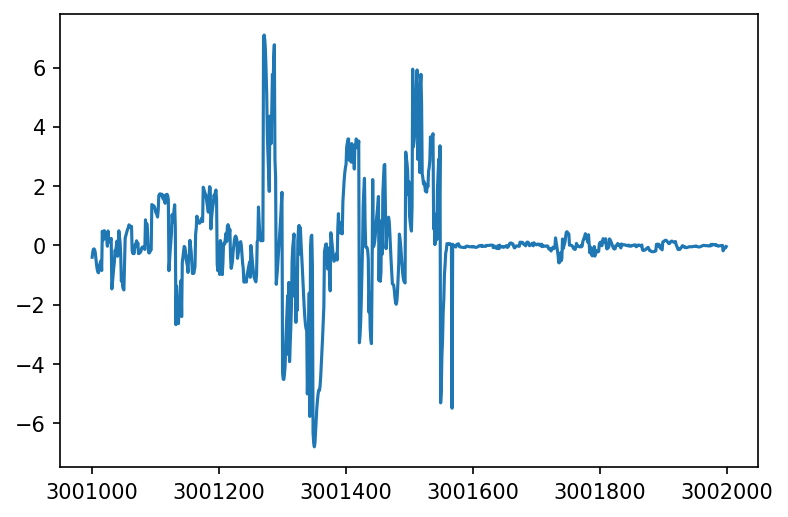}
    \caption{\textit{gyr}}
  \end{subfigure}
  \begin{subfigure}[b]{0.5\textwidth}
    \includegraphics[width=\textwidth]{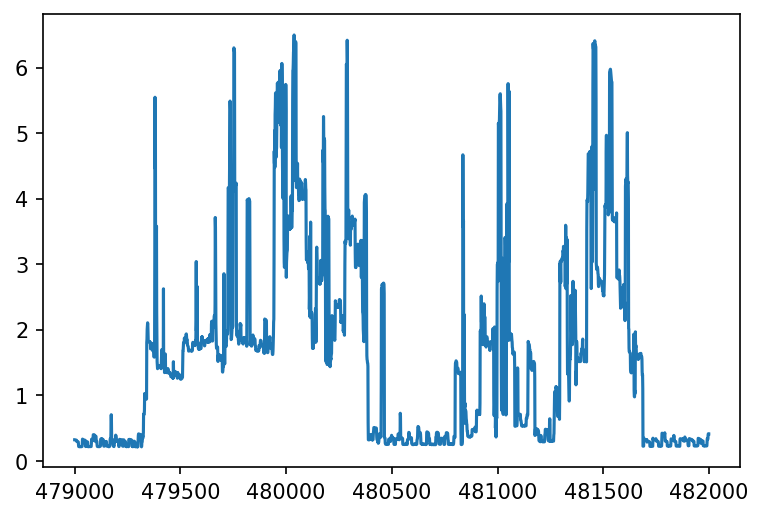}
    \caption{\textit{pow}}
  \end{subfigure}
  \begin{subfigure}[b]{0.5\textwidth}
    \includegraphics[width=\textwidth]{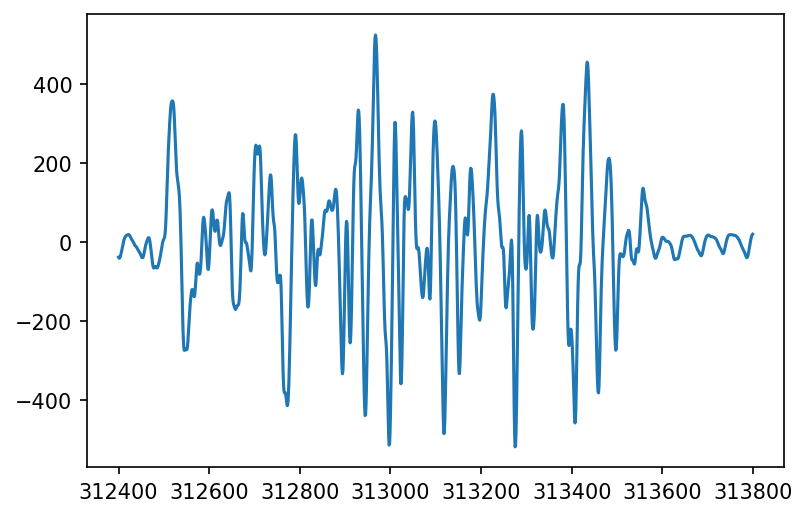}
    \caption{\textit{ppg}}
  \end{subfigure}
  \begin{subfigure}[b]{0.5\textwidth}
    \includegraphics[width=\textwidth]{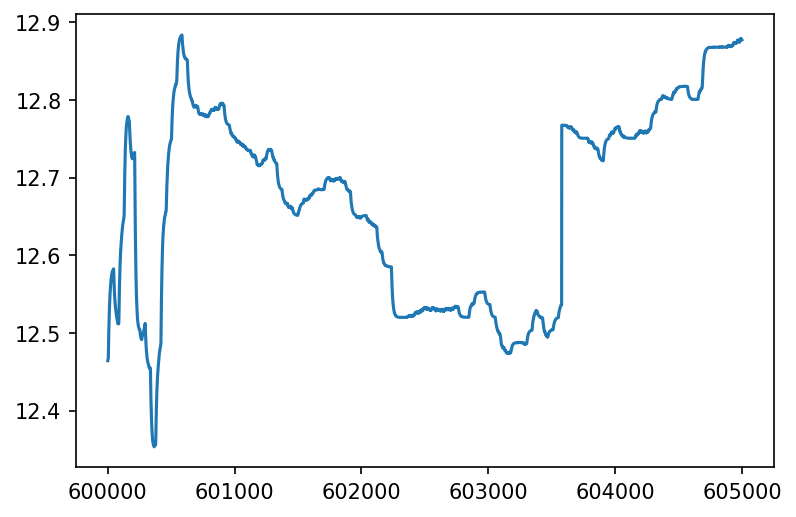}
    \caption{\textit{gas}}
  \end{subfigure}
  \begin{subfigure}[b]{0.5\textwidth}
    \includegraphics[width=\textwidth]{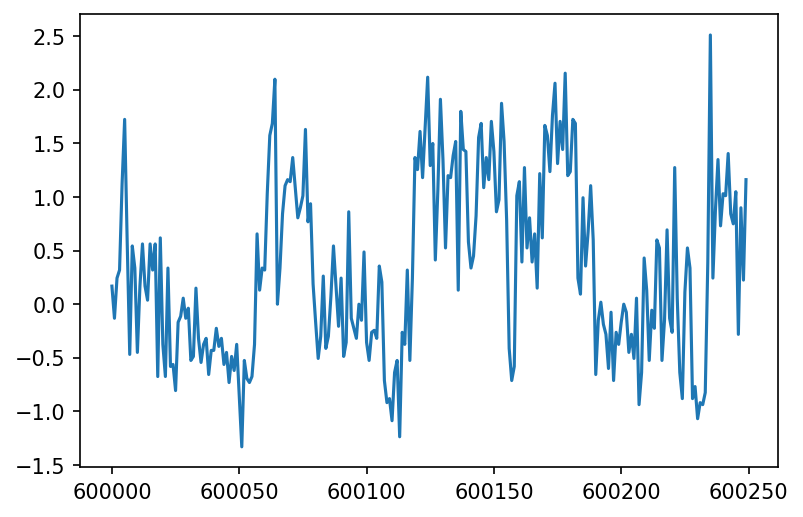}
    \caption{\textit{dna}}
  \end{subfigure}
  \begin{subfigure}[b]{0.5\textwidth}
    \includegraphics[width=\textwidth]{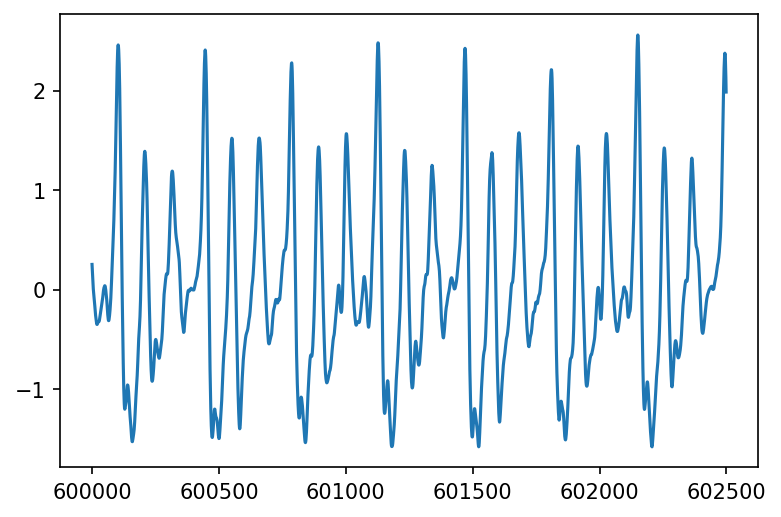}
    \caption{\textit{vib}}
  \end{subfigure}
  \begin{subfigure}[b]{0.5\textwidth}
    \includegraphics[width=\textwidth]{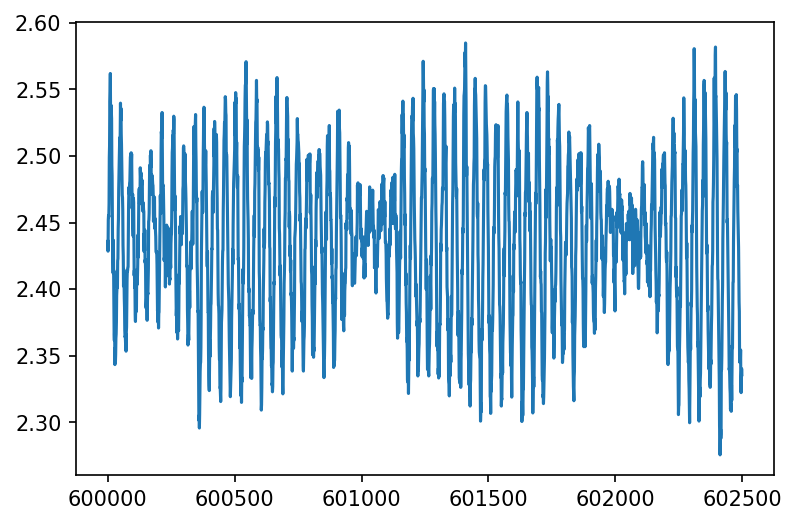}
    \caption{\textit{sen}}
  \end{subfigure}
	\caption{\textbf{Appendix:} Snapshots of the datasets used for evaluation.}
	\label{fig:plots}
\end{figure}
\end{document}